\documentclass[graybox]{svmult}

\usepackage{type1cm}   
\usepackage{makeidx}         % allows index generation
\usepackage{graphicx}        % standard LaTeX graphics tool
\usepackage{multicol}        % used for the two-column index
\usepackage[bottom]{footmisc}
\usepackage{amsmath}
\usepackage{amsfonts}
\usepackage{amssymb}
\newcommand{\nn}{\nonumber}
\usepackage{url}

\usepackage[utf8]{inputenc}
\usepackage{graphicx}
\usepackage{array}
\usepackage{float}
\usepackage{booktabs}
\usepackage{longtable}
\usepackage{subcaption}
\usepackage{multirow}
\usepackage{color}
\usepackage{caption}
\usepackage{subcaption}
\usepackage{slashed}
\usepackage{manyfoot}
\global\long\def\d{\partial}

\newcommand{\be}{\begin{equation}}
\newcommand{\ee}{\end{equation}}
\newcommand{\bea}{\begin{eqnarray}}
\newcommand{\eea}{\end{eqnarray}}

\def\s1{\hat s}

\begin{document}
\mainmatter              % start of a contribution
\title*{GeV scale dark matter and $B\rightarrow K+$ missing energy phenomenology in the $U(1)_{B-L}$ model }
\titlerunning{GeV scale dark matter and $B\rightarrow K+$ missing energy phenomenology}  % abbreviated title (for running head)

\author{Ajay Kumar Yadav, Shivaramakrishna Singirala,  Manas Kumar Mohapatra and Suchismita Sahoo}

\institute{Ajay Kumar Yadav \at Department of Physics, Central University of Karnataka, Kalaburagi-585367 \email{yadavajaykumar286@gmail.com}
\and Shivaramakrishna Singirala \at University of Hyderbad, Hyderabad, Telangana-500046 \email{krishnas542@gmail.com}\and Manas Kumar Mohapatra \at University of Hyderbad, Hyderabad, Telangana-500046 \email{manasmohapatra12@gmail.com} \and  Suchismita Sahoo \at Department of Physics, Central University of Karnataka, Kalaburagi-585367 \email{suchismita8792@gmail.com}}

%\institute{
%Department of Physics, Central University of Karnataka, Kalaburagi-585367, India \\ School of Physics, University of Hyderabad, Hyderabad-500046, India}
%
\authorrunning{Ajay Kumar Yadav et al.} % abbreviated author list (for running head)
%\tocauthor{Ivar Ekeland, Roger Temam, Jeffrey Dean, David Grove, Craig Chambers, Kim B. Bruce, and Elisa Bertino}
%
%\institute{School of Physics, University of Hyderabad, Hyderabad - 500046, India \\
%\email{suchismita@uohyd.ac.in$^*$,}
%\email{ rukmani98@gmail.com}}
%\\ WWW home page:
%\texttt{http://users/\homedir iekeland/web/welcome.html}
%\and
%Universit\'{e} de Paris-Sud,
%Laboratoire d'Analyse Num\'{e}rique, B\^{a}timent 425,\\
%F-91405 Orsay Cedex, France

\maketitle              % typeset the title of the contribution
\abstract{We elucidate the recently observed discrepancy in the branching ratio of $B^{+}\rightarrow K^{+}$ +inv decay mode using a GeV scale scalar dark matter in an anomaly-free $U(1)_{B-L}$ gauge extension of the Standard
Model.  We constrain the new parameters by using consistency with the existing bounds on Dark matter relic density, direct detection, collider and BR($B\rightarrow K^*\nu\bar{\nu}$) measured at the Belle II experiment. We investigate couplings between the mediator and the SM fermions as well as the dark matter particle. We then estimate the branching ratios of  $b \to s \nu \bar \nu$ decay processes such as $B\rightarrow (K^+,K^*)\nu\bar{\nu}$, $B_s\rightarrow (\eta,\eta')\nu\bar{\nu}$, $ B_s\rightarrow \phi\nu\bar{\nu}$ and $B_c\rightarrow (D_s, D_s^*)\nu\bar{\nu}$ in a common parameter space, meeting the current experimental bounds of both sectors simultaneously.}
%\newpage
\section{Introduction}
The extensively established foundational theory in particle physics, 
 Standard Model (SM) falls short in elucidating the asymmetry between matter and antimatter, the existence of dark matter (DM), and the observation of minute neutrino mass, which distinctly point towards the existence of new physics (NP) beyond the limitations of the SM. Over the recent years, noteworthy deviations of $(2-3)\sigma$ \cite{Capdevila:2023yhq, LHCb:2022qnv, LHCb:2017vlu, HFLAV:2022esi} from the Standard Model have surfaced in a variety of angular observables linked to flavor changing neutral current (FCNC) $b \to s l^+ l^-$ and flavor changing charge current (FCCC) $b \to c \tau \nu_l$ decay modes. The notable anomalies involve the lepton flavor universality violating (LFUV) ratios such as $R_{K^{(*)}}$, with disparities of $0.2\sigma$ \cite{LHCb:2022qnv}, $R_{D^{(*)}}$, with disparities of $2\sigma~(2.2 \sigma)$ \cite{HFLAV:2022esi} and $R_{J/\psi}$, with $2\sigma$ discrepancy \cite{LHCb:2017vlu}, in which the hadronic uncertainties cancelled out
significantly. In addition, the Belle II experiment\cite{Belle} has recently announced the initial measurement of $B^+\rightarrow K^+\nu\bar{\nu}$ branching ratio 
\begin{equation}
    \mathcal{BR}(B^{+}\rightarrow K^{+}\nu \bar{\nu})|^{\rm Expt} = \left(2.4\pm 0.7\right)\times 10^{-5} 
\end{equation}
which deviates from the SM prediction
\begin{equation}
    \mathcal{BR}(B^+\rightarrow K^{+}\nu\bar{\nu})|^{\rm SM} = (5.06\pm0.14\pm0.28)\times10^{-6}
\end{equation} 
 by $2.8\sigma$ \cite{Becirevic:2023aov}. This scenario requires a thorough examination of these discrepancies in both the SM and scenarios beyond the SM,  with the goal of investigating the framework of New Physics.

In $B \to K+ \nu_l \bar \nu_l$ decay, the undetected neutrinos carry away energy and momentum, contributing to the observed missing energy. Neutrinos escape detection due to their weak interaction, and any process with missing energy $E$ in the final state, such as $B^+ \rightarrow K^+ + \slashed{E}$, contributes to the measured branching ratio of the decay $B^+ \rightarrow K^{+}\nu\bar{\nu}$. The Standard Model requires supplementation with extra symmetries or particles to address this anomaly in missing energy. The missing energy can be thought of as dark matter pair, a pair of sterile neutrino or any other light particle participating in this $b \to s  + \slashed{E}$ transition with a weak enough interaction strength. Therefore, precision investigations into B meson decay are vital to identify or constrain Dark Matter or other light Beyond SM elements. The combined study of Dark matter and NP suggestion from B meson decay is an exciting window towards the building of new physics. In this work, the SM is extended with $U(1)_{B-L}$ gauge extensions to fix the missing energy anomaly with a light dark matter. Furthermore, we compute the branching ratios of other  $b \to s \nu \bar \nu$ decay modes such as $B\rightarrow K^{(*)}\nu\bar{\nu}$, $B_s\rightarrow \eta^{(')} \nu\bar{\nu}$,$ B_s\rightarrow \phi\nu\bar{\nu}$ and $B_c\rightarrow D_s^{(*)}\nu\bar{\nu}$ to recheck the real existence of NP. 

The paper is organised as follows. Section 2 describes the particle content and the interaction Lagrangian of the gauge extended model. We present the constraints on new parameters from both the dark and flavor sectors in section 3.  Section 3 also includes the general effective Hamiltonian for $b \to s \nu_l \bar \nu_l$ channels. The impacts of new parameters on $B\rightarrow P(V)\nu_l \bar{\nu_l}$ decay modes are discussed in section 4. Section 5 outlines the concluding remarks.

\section{Model Description}
Within the framework of $U(1)_{B-L}$ gauge extensions of the Standard Model, we illustrate a fundamental case involving singlet scalar dark matter to explain the absence of energy as dark matter.  For anomaly cancellation, we strengthen the model by incorporating the three Majorana right-handed neutrinos $N_{iR}$ with $i=1,2,3$, each carrying a charge $-1$ under the extended $U(1)_{B-L}$ gauge group. Additionally, we include two singlet scalars, $\phi$ (spontaneously breaks the new $U(1)$) and $\phi_{\rm DM}$ (possible dark matter candidate) with $B-L$ charges $2$ and $1/3$ respectively as given in the  Table: \ref{tab:BL}.
%================================================
\begin{table}[htb]
\begin{center}
%================================================
\begin{tabular}{|c|c|c|c|}
	\hline
			& Field	& $ SU(2)_L\times U(1)_Y$	& $U(1)_{B-L}$	\\
	\hline
	\hline
	Fermions	& $Q_L \equiv(u, d)^T_L$			& $(\textbf{2},~ 1/6)$	& $1/3$	\\
			& $u_R$							& $(\textbf{1},~ 2/3)$	& $1/3$	\\
			& $d_R$							& $(\textbf{1},~-1/3)$	& $1/3$	\\
			& $\ell_L \equiv(\nu,~e)^T_L$	& $(\textbf{2},~  -1/2)$	&  $-1$	\\
			& $e_R$							& $(\textbf{1},~  -1)$	&  $-1$	\\
			& $N_{1R}, N_{2R}, N_{3R}$						& $(\textbf{1},~   0)$	&  $-1$	\\
	\hline
	Scalars	& $H$							& $(\textbf{2},~ 1/2)$	&   $0$	\\
			& $\phi$						& $(\textbf{1},~   0)$	&  $2$	\\  
			& $\phi_{\rm DM}$						& $(\textbf{1},~   0)$	&  $1/3$	\\  
	\hline
	\hline
\end{tabular}
%================================================
\caption{Fields and their charges of the proposed $U(1)_{B-L}$ model.}
\label{tab:BL}
\end{center}
\end{table}

The pertinent interaction Lagrangian is given as follows:
\begin{eqnarray}
\label{eq:TheModel}
	\mathcal{L}_\text{BL} 
	&&= -\frac{1}{3} \,g_\text{BL}\overline{Q}_L \,Z_\mu^\prime \gamma^\mu Q_L   
            -\frac{1}{3}\,g_\text{BL} \overline{u}_R \,Z_\mu^\prime \gamma^\mu u_R
            -\frac{1}{3} \,g_\text{BL}\overline{d}_R \,Z_\mu^\prime \gamma^\mu d_R + \,g_\text{BL}\overline{\ell}_L \,Z_\mu^\prime \gamma^\mu \ell_L
            + \,g_\text{BL} \overline{e}_R \,Z_\mu^\prime \gamma^\mu e_R  \nonumber \\
        &&~~~       + i \, \overline{N}_{iR} \left( \slashed{\d} + i\,g_\text{BL} \,Z_\mu^\prime \gamma^\mu \right)\,N_{iR} - \frac{y_{\alpha \beta}}{2} \left( \sum_{\alpha, \beta = 1}^{3}\overline{N^c_{\alpha R}}\,N_{\beta R} \,\phi + h.c.\right)  \nonumber \\ 
	&&~~~ + \left|\left( \d_\mu - 2ig_\text{BL}Z'_\mu \right) \phi \right|^2 + \left|\left( \d_\mu - \frac{1}{2}ig_\text{BL}\,Z'_\mu \right) \phi_{\rm DM}\right|^2 	     - V \left( H, \phi,\phi_{\rm DM}\right)
	     + \mathcal{L}_\text{SM} 
	  \, ,
\end{eqnarray}
where the scalar potential $V(H,\phi,\phi_{\rm DM})$ of the model is given by
 \begin{align}
V(H,\phi,\phi_{\rm DM}) &= \mu^2_H  H^\dagger H + \lambda_H (H^\dagger H)^2  + \mu^2_{\phi} \phi^\dagger \phi + \lambda_{\phi} (\phi^\dagger \phi)^2  +\mu^2_{\rm DM} \phi^\dagger_{\rm DM} \phi_{\rm DM}  \nonumber \\ 
& + \lambda_{\rm DM} (\phi^\dagger_{\rm DM} \phi_{\rm DM})^2 + \lambda_{H\phi} (H^\dagger H) (\phi^\dagger \phi) +\lambda_{\rm HD} (H^\dagger H) (\phi^\dagger_{\rm DM} \phi_{\rm DM})\nonumber \\  &  + \lambda_{{\rm D}\phi} (\phi^\dagger \phi) (\phi^\dagger_{\rm DM} \phi_{\rm DM}). \label{scalarpot}
\end{align} 
with $\phi_{\rm DM} = \frac{S+iA}{\sqrt{2}}$, $\langle H\rangle=(0,v/\sqrt2)^T$, $\langle \phi_2\rangle=v_2/\sqrt2$. After spontaneous symmetry breakdown, $\phi$ attains a vev $v_2$ and the new gauge field $Z^\prime$ receives a mass $M_{Z^\prime} = 2 g_{\rm BL} v_2$, where $g_{\rm BL}$ is new gauge coupling. The mixing of the two CP-even scalars, $H_1$ and $\phi$, results in two mass eigenstates: $H_1$ (the observed 125 GeV Higgs at LHC) and a lighter scalar $H_2$ (with a mass below 5 GeV).

\section{ Constraints of New Parameters}
After acquiring an understanding of the new particles and their interactions,  we now proceed to constrain the new parameters from both the flavor phenomenology and the dark matter sectors. 
\subsection{Dark Matter Phenomenology}
%\begin{itemize}
%    \item \textbf{Relic Density}:
\subsubsection{Relic Density}
    \begin{figure}[htb]
\centering
\includegraphics[width=0.45\linewidth]{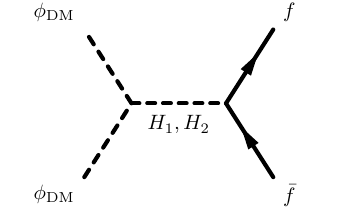}
\includegraphics[width=0.45\linewidth]{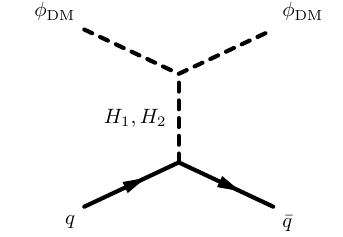}
\caption{Feynman diagrams contributing to relic density (left panel) and direct detection (right panel). }
\label{feyn}
\end{figure}
\begin{figure}[htb]
\centering
\includegraphics[width=0.6\linewidth]{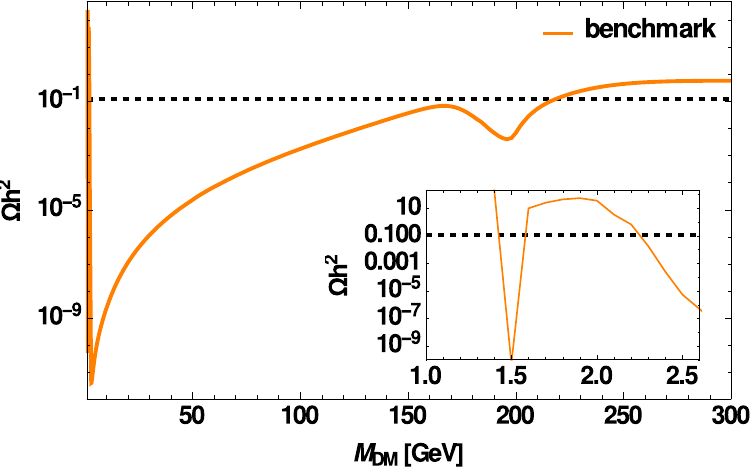}
\caption{Relic density vs DM mass with $M_{H_2} = 3$ GeV. Black horizontal dotted lines denote the $3\sigma$ range of Planck limit \cite{Planck:2015fie}.}
\label{relicplot}
\end{figure}

Both components of the singlet scalar have the potential to contribute to the relic density of the Universe. The dark matter relic abundance is calculated using 
\begin{equation}
\label{eq:relicdensity}
\Omega \text{h}^2 = \frac{1.09 \times 10^{9} ~{\rm{GeV}}^{-1}}{ {g_\ast}^{1/2} M_{\rm{pl}}}\frac{1}{J(x_f)},
\end{equation}
where $M_{\rm{pl}}=1.22 \times 10^{19} ~\rm{GeV}$ is the Planck mass, $g_\ast = 106.75$ denotes the total number of effective relativistic degrees of freedom,  and $J(x_f)$  is constructed as 
\begin{equation}
J(x_f)=\int_{x_f}^{\infty} \frac{ \langle \sigma v \rangle (x)}{x^2} dx.
\end{equation}
with, $\langle \sigma v \rangle$ is the thermally averaged annihilation cross-section \cite{Singirala:2017see}. The scalar dark matter (DM) can undergo annihilation through scalar bosons $H_{1,2}$, or $Z^\prime$. We focus on the scalar portal to achieve a light DM ($M_{DM} < 2.5$ GeV), aiming to account for the missing energy signal observed in B-meson decays. The annihilation channels encompass $SS (AA) \to f\bar{f}$, where $f$ represents quark-antiquark and lepton-antilepton pairs, constituting s-channel processes (left panel of Fig. \ref{feyn}). Consequently, a resonance is evident when the DM mass is approximately half of the propagator mass, $M_{\rm DM} \approx \frac{M_{H_2}}{2}$,  in the plot (Fig. \ref{relicplot}) of $\Omega h^2$ versus DM mass. We have implemented the model in LanHEP \cite{Semenov:1996es} and then micrOMEGAs \cite{Belanger:2006is} packages for dark matter study. 
%\item \textbf{Direct Detection}:
\subsubsection{Direct Detection}
\begin{figure}[htb]
\centering
\includegraphics[width=0.6\linewidth]{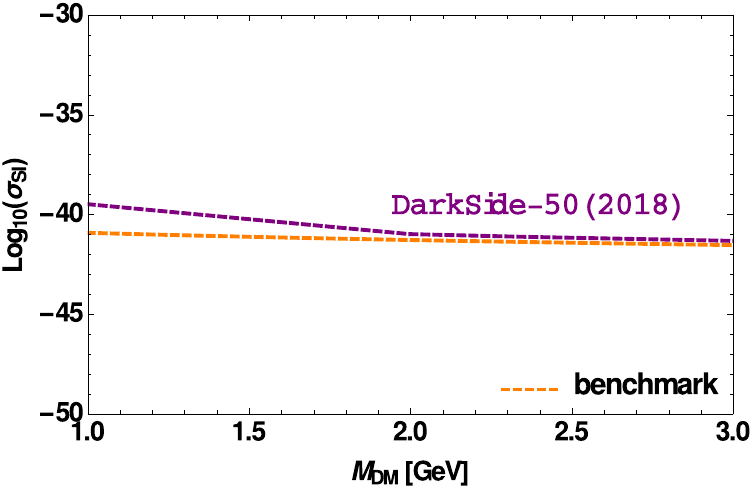}
\caption{WIMP-nucleon spin-independent cross section vs DM mass. Here the dashed line corresponds to DarkSise-50 \cite{DarkSide:2018kuk}.}
\label{DDplot}
\end{figure}

The dark matter can disperse off the nucleus through the effective interaction of the scalars $H_1$ and $H_2$. The DM-nucleon effective Lagrangian is given as

\begin{equation}
    \mathcal{L}_{\rm eff} = \frac{\lambda}{M^2_{H_i}} SS \bar{q}q + \frac{\lambda}{M^2_{H_i}} AA  \bar{q}q.
\end{equation}
The spin-independent (SI) cross section vs DM mass is projected in Fig. \ref{DDplot} where the cross-section is discovered to be significantly less than DarkSide-50's experimental upper limit \cite{DarkSide:2018kuk}.
In case of $Z^\prime$-portal, the effective Lagrangian is follows as
\bea%
\mathcal{L^\prime}_{\mathrm{eff}}\supset
-\frac{{n_{\rm DM}} g_{\rm BL}^2}{3 M_{Z^{\prime}}^2}
\left(S\partial^{\mu}A-A\partial^{\mu}S  \right)\bar{u}\gamma_\mu u
-\frac{{n_{\rm DM}}g_{\rm BL}^2}{3 M_{Z^{\prime}}^2}
\left(S\partial^{\mu}A-A\partial^{\mu}S  \right) \bar{d}\gamma_\mu d,\nn\\%
\eea
where $n_{\rm DM}$ is the charge of DM under new $U(1)$. The expression for SI WIMP-nucleon cross section is given by \cite{Khalil:2011tb,Singirala:2017see}
\be
\sigma_{Z^{\prime}}=\frac{{\mu}^2}{16\pi}\frac{{n^2_{\rm DM}} g_{\rm BL}^4}{M_{Z^{\prime}}^4}.
\ee
where $\mu = \left(\frac{M_n M_{\rm DM}}{M_n + M_{\rm DM}}\right)$ is the reduced mass of DM-nucleon with $M_n$ being the nucleon mass. It is convenient to write (in $\text{cm}^2$) as
\begin{equation}
 \sigma_{Z^{\prime}}  = 7.75 \times 10^{-42} \times {n_{\rm DM}}^2 \times \left( \frac{\mu}{1 ~{\rm{GeV}}}\right)^2  \times \left( \frac{1 ~{\rm{TeV}}}{\left(\frac{M_{Z^{\prime}}}{g_{\rm BL}}\right)}\right)^4 .\hspace{2.5 truecm}
\end{equation}
From the above expression, one can see that the model can produce SI contribution below DarkSide-50 limit as well. We truncate discussion in gauge portal to this point as our emphasis is on scalar-portal dark matter for this work.

%\item \textbf{Higgs Invisible Width}: 
\subsubsection{Higgs Invisible Width}
Since Higgs can decay to light scalars $(S, A, H_2)$, constraints on Higgs invisible width may be able to limit the scalar couplings. The effective Lagrangian with scalar $H_2$ is given as
\begin{equation}
    \Gamma_{\rm inv} = \Gamma_{H_2 H_2} + \Gamma_{SS} + \Gamma_{AA}.
\end{equation}

We utilize the upper limit on Higgs invisible width, i.e., $20\%$ of total width in Fig. \ref{Hdecay}
\begin{figure}[htb]
\centering
\includegraphics[width=0.6\linewidth]{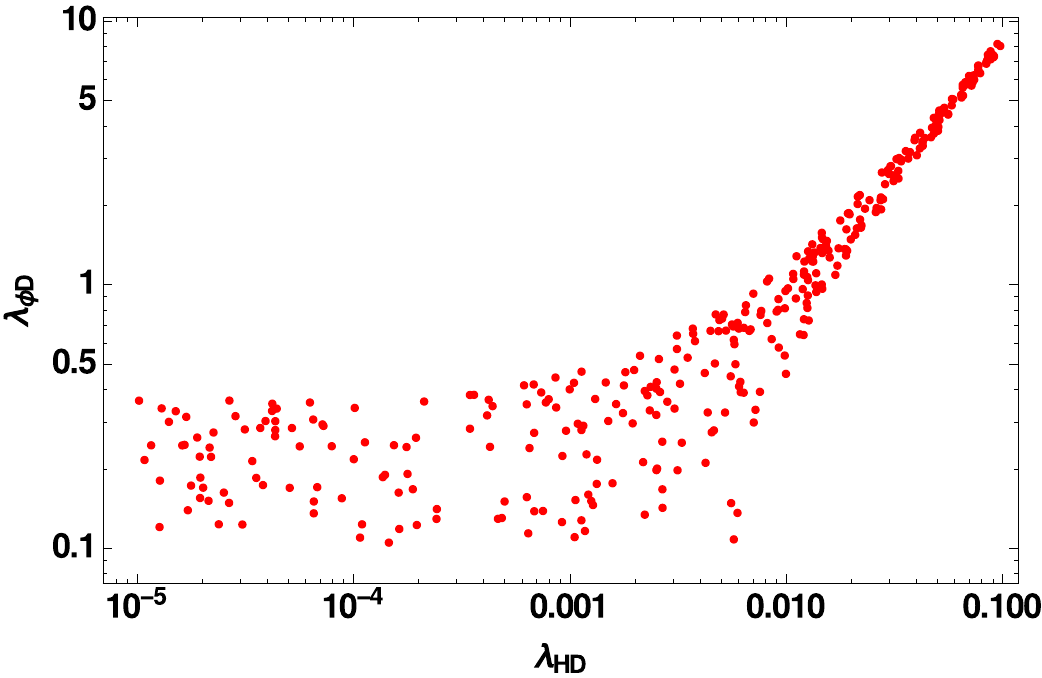}
\caption{Constraint on scalar couplings from Higgs invisible width.}
\label{Hdecay}
\end{figure}
%\end{itemize}
\subsection{Flavor Phenomenology: Constraints from $B \to K+$ missing energy}
%\subsubsection{Effective Hamiltonian}
The general  effective Hamiltonian of $b\rightarrow s\nu\bar{\nu}$ quark level transitions at the scale of the bottom quark mass is stated as follows \cite{Altmannshofer:2009ma, Buras:2014fpa} 
\begin{equation}
    \mathcal{H}_{eff}=\frac{-4 G_F}{\sqrt{2}} V_{tb}V_{ts}^{*} (C_R^{\nu} O_R^{\nu}+C_R^{\nu} O_R^{\nu})+ h.c. ,
\end{equation}
where $G_F$ is the Fermi constant, $V_{qq^\prime}$ are the CKM elements,  $\mathcal{O}_{L,R}^{\nu}$ are the  six-dimension operators 
\begin{equation}
    O_L^{\nu}=\frac{\alpha^2}{4\pi}(\overline{s}\gamma_{\mu}P_L b)(\overline{\nu}\gamma^{\mu}(1-\gamma_5)\nu),~~  \\
    O_R^{\nu}=\frac{\alpha^2}{4\pi}(\overline{s}\gamma_{\mu}P_R b)(\overline{\nu}\gamma^{\mu}(1-\gamma_5)\nu)
\end{equation}
and $C_{L,R}^{\nu}$ are the corresponding Wilson coefficients. Here, $C_L^{\nu}=-6.38\pm 0.06$ and $C_R^{\nu}=0$ in the SM, which can be nonzero in the presence of new physics. 

In $B\to K+$ Missing Energy, the disagreement between theoretical and experimental branching ratio $\Longrightarrow$
The presence of a new source of missing energy $\Longrightarrow$ can be a pair of DM.
\begin{equation}
    \mathcal{BR}(B\rightarrow K+\slashed{E})=\mathcal{BR}(B\rightarrow K\nu\bar{\nu}) + \mathcal{BR}(B\rightarrow K\phi_{\rm{DM}}\phi_{\rm{DM}}). 
\end{equation}
Here the  branching fraction for $B\rightarrow K\nu\bar{\nu}$ decays with respect to $q^2$ is given by \cite{Colangelo:1996ay} 
\begin{equation}
   \frac{d\mathcal{BR}(B\rightarrow K \nu\bar{\nu})}{d q^2} =\frac{G_F^2 \alpha^2}{256 \pi^5 m_B^3}\tau_B\mid V_{tb}V_{ts}^*\mid^2\lambda^{3/2}(m_B^2,m_K^2,q^2)[f_+(q^2)]^2\mid C_L^{\nu} + C_R^{\nu} \mid^2,\label{abc}
\end{equation}
with $m_B (m_K)$ are the masses of $B(K)$ meson and $f_+(q^2)$ form factors are taken from the Light Cone Sum Rules (LCSRs) method \cite{Gubernari:2018wyi}. All possible one loop penguin diagrams contributing to the $B\rightarrow K\phi_{\rm{DM}}\phi_{\rm{DM}}$ decay modes are presented in the Fig. \ref{Hdecay}. 
\begin{figure}[htb]
\centering
\includegraphics[width=0.4\linewidth]{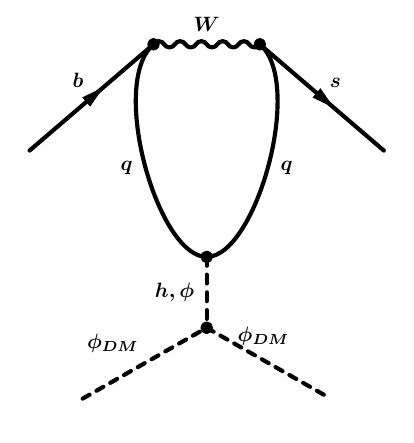}
\includegraphics[width=0.4\linewidth]{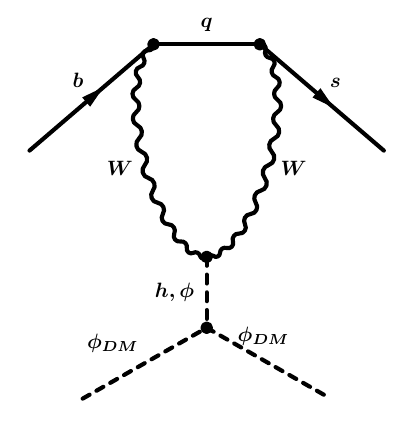}
\caption{Feynman diagrams contributing to missing energy}
\label{Hdecay}
\end{figure}

\section{Impact on $B\rightarrow P(V)\nu\bar{\nu}$ Decay Modes}
This section discusses the implication of the constrained new parameters on the other $b \to s \nu_l \bar \nu_l$ decay modes. The branching ratio expression for $B \to P \nu \bar \nu_l$ decay modes, where $P$ is a pseudo scalar meson, is given in the previous section.

For the study of $B\rightarrow V\nu\bar{\nu}$ decay modes, where $V$ denotes for vector meson, the branching fraction is given as\cite{Buras:2014fpa}:
\begin{equation}
    \frac{d\mathcal{BR}(B\rightarrow V\nu\bar{\nu})}{d q^2}=3 \tau_B[|A_{\perp}|^2 + |A_{\parallel}|^2 +|A_0|^2]. 
\end{equation}
Here, $A_{\perp,\parallel,0}$ are the $B\rightarrow V$ transversity amplitudes, which are expressed in terms of WCs and form factors as 
\begin{align}
A_{\perp}(q^2) &= \frac{2N\sqrt{2\lambda(m_B^2,m_P^2,q^2)}}{m_B}[C_L^{\nu}+C_R^{\nu}]\frac{V(q^2)}{[m_V+m_B]}\\ 
A_{\parallel}(q^2) &= \frac{-2N\sqrt{2}}{m_B}[m_V+m_B][C_L^{\nu}-C_R^{\nu}]A_1(q^2)\\ 
A_0(q^2) &= \frac{-N m_B}{m_V \sqrt{q^2}}[C_L^{\nu}-C_R^{\nu}]\Big(\Big[1-\frac{m_V^2}{m_B^2}-\frac{q^2}{m_B^2}\Big][m_V+m_B]A_1(q^2)\\
& -\lambda(m_B^2,m_P^2,q^2)\frac{A_2(q^2)}{[m_B+m_V]}\Big)
\end{align}
where $N$ is the normalization factor 
\begin{equation*}
    N=|V_{tb}V_{ts}^*|\Big[\frac{G_F^2 \alpha^2 q^2 \sqrt{\lambda(m_B^2,m_P^2,q^2)}}{3.2^10 \pi^5 m_B}\Big]^{1/2}
\end{equation*} 
with $\lambda(m_B^2,m_V^2,q^2)=m_B^4 +m_V^4 +q^4- 2( m_B^2 m_V^2 +m_V^2 q^2 + m_B^2 q^2 )$.

For numerical analysis, we use the particle masses, lifetime of $B_{(s)}$ meson, CKM matrix elements and Fermi constant values from the Particle Data Group \cite{ParticleDataGroup:2022pth}. We use the form factor results that either permit comparison with the experiment or enable us to make predictions. For $B^+\rightarrow K^+ \nu\bar{\nu}$, we used form factor fit result of LCSRs  with LQCD\cite{Gubernari:2018wyi, Parrott:2022rgu, Grunwald:2023nli}. The $B\rightarrow K^*, \phi$ form factors, computed in the LCSR method, which is used from  Ref.\cite{Bharucha:2015bzk}. The form factor for $B\rightarrow\eta,\eta'$ are taken from Ref.\cite{Duplancic:2015zna, Ball:2004ye} which are obtained in the LCSRs approach and the form factors for $B_c\rightarrow (D_s, D_s^*)\nu\bar{\nu}$ in the  QCD and RQM method respectively are used from Ref.\cite{Cooper:2021bkt, Mohapatra:2021ynn}. The predicted branching ratios of $B \to P(V) \nu_l \bar \nu_l$ decay modes in the SM and in the extended $U(1)_{B-L}$ model are presented in the Table \ref{tab:my_label}. 
\begin{table}[htb]
    \centering
\begin{tabular}{|c|c|c|c|}
    \hline
        \textbf{Decay Modes} &   \textbf{Values in SM} &   \textbf{Values in $U(1)_{B-L}$ Model} & \textbf{Experimental limit} \\ \hline \hline
 %       $B^+\rightarrow K^+ \nu \bar{\nu}$& $1.185 \times 10^{-6}$ & - & $2.4\times10^{-5}$ \\
         $B^0\rightarrow K^{*0} \nu \bar{\nu}$& $1.6285 \times 10^{-6}$ & $3.91 \times 10^{-6}$ & $< 1.8\times 10^{-5} $\\
 %        $B^+\rightarrow K^{*+} \nu \bar{\nu}$& $2.1172 \times 10^{-6}$ & --& ---\\
         $B_s\rightarrow \phi \nu \bar{\nu}$& $2.262\times 10^{-6}$ & $5.423 \times 10^{-6}$ & $< 5.4\times 10^{-3}$ \\
          $B \rightarrow \eta \nu \bar{\nu}$ & $0.61\times10^{-6}$ & $1.47 \times 10^{-6}$ & $\cdots$ \\
         $B \rightarrow \eta' \nu \bar{\nu}$ & $0.37\times10^{-6}$ & $0.97 \times 10^{-6}$ & $\cdots$ \\
         $B_c^+\rightarrow D_s^+ \nu \bar{\nu}$ & $0.185 \times 10^{-6}$& $0.52 \times 10^{-6}$ & $\cdots$ \\
        $B_c^+\rightarrow D_s^{*+} \nu \bar{\nu}$ & $0.086 \times 10^{-6}$& $0.215 \times 10^{-6}$ & $\cdots$ \\

         \hline
    \end{tabular}
    \caption{Predicted branching ratios of $b \to s \nu \bar \nu$ channels in the proposed $U(1)_{B-L}$ model. }
    \label{tab:my_label}
\end{table}

\section{Conclusion}
In this paper, we have discussed the missing energy anomaly associated with $b\rightarrow s\nu_l\bar{\nu_l}$ channel in the context of an anomaly-free $U(1)_{B-L}$ gauge extended model. We addressed the $B\rightarrow K+$ missing energy anomaly as the presence of $GeV$ scale scalar dark matter pair. We constrained the new parameters by using the existing experimental bounds on the dark matter and flavor physics phenomenology. We then predicted the branching ratios of other $b\rightarrow s\nu_l\bar{\nu_l}$ decay modes by using the constrained new parameters,  which are found to be within reach of the experimental limit.

\section*{Acknowledgement}
AKY would like to thank the Department of Science and Technology (DST)- Inspire Fellowship division, Govt of India, for the financial support through ID No. IF210687.

%\section{References}
\bibliographystyle{unsrt}
\bibliography{main}
%\bibliographystyle{apsrev4-1}
%\bibliography{reference}

\end{document}